\catcode`\@=11

\newif\if@fewtab\@fewtabtrue
{\count255=\time\divide\count255 by 60
\xdef\hourmin{\number\count255}
\multiply\count255 by-60\advance\count255 by\time
\xdef\hourmin{\hourmin:\ifnum\count255<10 0\fi\the\count255}}
\def\ps@draft{\let\@mkboth\@gobbletwo
    \def\@oddfoot{\hbox to 7 cm{\tiny \versionno
       \hfil}\hskip -7cm\hfil\rm\thepage \hfil {\tiny\draftdate}}
    \def\@oddhead{}
    \def\@evenhead{}\let\@evenfoot\@oddfoot}
\def\draftdate{\number\month/\number\day/\number\year\ \ \ \hourmin }

\global\def\draftcontrol{0}

\def\draftcite#1{\ifnum\draftcontrol=1#1\else{}\fi}
\def\@lbibitem[#1]#2{\item{}\hskip -3\hbox to 2cm
{\hfil$\scriptstyle\draftcite{#2}$}\hskip
1cm[\@biblabel{#1}]\if@filesw
     {\def\protect##1{\string ##1\space}\immediate
      \write\@auxout{\string\bibcite{#2}{#1}}}\fi\ignorespaces}

\def\@bibitem#1{\item\hskip -3cm \hbox to 2cm
{\hfil {\footnotesize\draftcite{#1}}}\hskip 1cm
\if@filesw \immediate\write\@auxout
       {\string\bibcite{#1}{\the\value{\@listctr}}}\fi\ignorespaces}

\def\citen#1{\if@filesw \immediate\write \@auxout {\string\citation{#1}}\fi%
\@tempcntb\m@ne \let\@h@ld\relax \def\@citea{}%
\@for \@citeb:=#1\do {\@ifundefined {b@\@citeb}%
    {\@h@ld\@citea\@tempcntb\m@ne{\bf ?}%
    \@warning {Citation `\@citeb ' on page \thepage \space undefined}}%
    {\@tempcnta\@tempcntb \advance\@tempcnta\@ne
    \setbox\z@\hbox\bgroup\ifcat0\csname b@\@citeb \endcsname \relax
    \egroup \@tempcntb\number\csname b@\@citeb \endcsname \relax
    \else \egroup \@tempcntb\m@ne \fi \ifnum\@tempcnta=\@tempcntb
    \ifx\@h@ld\relax \edef \@h@ld{\@citea\csname b@\@citeb\endcsname}%
    \else \edef\@h@ld{\hbox{--}\penalty\@highpenalty
    \csname b@\@citeb\endcsname}\fi
    \else \@h@ld\@citea\csname b@\@citeb \endcsname \let\@h@ld\relax \fi}%
\def\@citea{,\penalty\@highpenalty\hskip.13em plus.13em minus.13em}}\@h@ld}
\def\@citex[#1]#2{\@cite{\citen{#2}}{#1}}%
\def\@cite#1#2{\leavevmode\unskip\ifnum\lastpenalty=\z@\penalty\@highpenalty\fi%
  \ [{\multiply\@highpenalty 3 #1%
  \if@tempswa,\penalty\@highpenalty\ #2\fi}]}   %
\makeatother 

\catcode`\@=12

\def\bound        {{\rm bound}}
\def\cala  {{\cal A}}
\def\calap {\bar{\cal A}}
\def\calh  {{\cal H}}

\def\cft           {conformal field theory}
\def\cfts          {conformal field theories}
\def\complex       {{\dl C}}
\def\dl            {\mathbb }
\def\Hom           {{\rm Hom}}
\newcommand\hsp[1] {\mbox{\hspace{#1 em}}}
\def\ii            {{\rm i}}

\newcommand\nxt[1] {\\\raisebox{.12em}{\rule{.35em}{.35em}}\hsp{.6}#1}

\def\reals         {{\dl R}}
\def\zet           {{\dl Z}}

\documentclass[12pt]{article} \usepackage{amssymb,amsfonts,latexsym}
\usepackage[dvips]{graphics} \usepackage{epsfig}
\begin{document}


\begin{flushright}  {~} \\[-1cm]
{\sf hep-th/0009111}\\{\sf PAR-LPTHE 00-33}
\\[1mm]
{\sf September 2000} \end{flushright}

\begin{center} \vskip 14mm
{\Large\bf SOLITONIC SECTORS, CONFORMAL BOUNDARY } \\[4mm]
{\Large\bf CONDITIONS AND THREE-DIMENSIONAL} \\[4mm]
{\Large\bf TOPOLOGICAL FIELD THEORY}
     $^\dagger$
\\[20mm]
{\large Christoph Schweigert$\;^1$ \ and \ J\"urgen Fuchs$\;^2$}
\\[8mm]
$^1\;$ LPTHE, Universit\'e Paris VI~~~{}\\
4 place Jussieu\\ F\,--\,75\,252\, Paris\, Cedex 05\\[5mm]
$^2\;$ Institutionen f\"or fysik~~~~{}\\
Universitetsgatan 1\\ S\,--\,651\,88\, Karlstad\\[5mm]
\end{center}
\vskip 18mm
\begin{quote}{\bf Abstract}\\[1mm]
The correlation functions of a two-dimensional rational conformal
field theory, for an arbitrary number of bulk and boundary fields and
arbitrary world sheets, can be expressed in terms of Wilson graphs in
appropriate three-manifolds. We present a systematic approach to boundary
conditions that break bulk symmetries. It is based on the construction,
by `$\alpha$-induction', of a fusion ring for the boundary fields.
Its structure constants are the annulus coefficients and its $6j$-symbols give
the OPE of boundary fields. Symmetry breaking boundary conditions correspond 
to solitonic sectors.
\end{quote}
  \vfill\noindent------------------\\
  $^\dagger$~Invited talk by Christoph Schweigert at the TMR conference
  ``Non-perturbative quantum effects 2000'', Paris, September 2000;
  to appear in the Proceedings.

\newpage 

Conformal field theory in two dimensions plays a fundamental role in the theory
of two-dimen\-si\-o\-nal critical systems of classical statistical mechanics
\cite{Card}, in quasi one-dimensional condensed matter physics \cite{Affl}
and in string theory \cite{POlc}. The study of defects in systems of
condensed matter physics \cite{osaf2}, of percolation probabilities 
\cite{card12} and of (open) string perturbation theory in the background of 
certain string solitons, the so-called D-branes \cite{polc3}, forces one to 
analyze conformal field theories on surfaces that may have boundaries 
and\,/\,or can be non-orientable.

In this contribution, we present a systematic description of correlation 
functions of an arbitrary number of bulk and boundary fields on general 
surfaces. It is based on the fundamental fact \cite{witt27,frki2}
that conformal blocks appear in two
different contexts: They are building blocks for the correlators of
two-dimensional \cfts, and they are the spaces of physical states
in topological field theories, TFT, in three dimensions.

For simplicity, we take the modular invariant torus partition function that 
encodes the spectrum of bulk fields of the theory to be of charge conjugation 
form, i.e.\ $Z_{\lambda,\mu}^{}\,{=}\,\delta_{\lambda,\mu^+_{\phantom i}}$.
We will, however, include in our discussion boundary conditions 
that do not preserve all bulk symmetries. We consider general cases of
symmetry breaking by boundary conditions. In particular, we do not have to 
require that left movers and right movers are linked, at the boundary, by some 
automorphism of the chiral algebra. Put differently, the subalgebra of
chiral symmetries that is preserved by the boundary conditions is not
necessarily an orbifold subalgebra. Applications of the theory include
non-BPS branes in interacting backgrounds and boundary conditions for
exceptional modular invariants.

We will start with a brief review of TFT in three dimensions, and then
formulate the basic problem that arises when one constructs a full
two-dimensional CFT from a chiral CFT. The amplitudes in the presence of
symmetry preserving boundary conditions will be discussed in Section 3.
Symmetry breaking boundary conditions are the subject of Section 4.

\section{Three-dimensional TFT}

The basic feature of three-dimensional TFT is that it provides a {\em modular 
functor\/}: To geometric data it associates algebraic structures. Concretely, 
it associates vector spaces -- the spaces $\calh(\hat X)$ of conformal blocks -- 
to two-dimensional manifolds $\hat X$, and to three-manifolds, endowed with
somewhat more structure, it assigns an endomorphism of these vector spaces. 

In thoses cases where the TFT can be defined in terms of path integrals, 
e.g.\ for Chern-Simons theories, the reader is invited to think of the vector 
space $\calh(\hat X)$  as the space of (gauge equivalence classes of)
boundary conditions for the fields appearing in the path integral and to 
think of the endomorphisms as transition amplitudes. We would like to stress, 
however, that our approach does not rely on
the existence of a path integral description.
In fact, the only necessary input is the structure of a modular
tensor category \cite{TUra}, which is a formalization of Moore-Seiberg data
like fusing and braiding matrices and conformal weights.

More precisely, conformal blocks are associated to {\em extended
surfaces\/}: These are two-dimen\-si\-o\-nal, oriented manifolds with a finite
collection of small arcs. Each arc carries a label from a set $I$.
In our application, these are primary fields, or equivalently, irreducible 
representations of a chiral symmetry algebra. Moreover, it is necessary
to choose a Lagrangian subspace of $H_1(\hat X,\reals)$. We will sometimes 
suppress these auxiliary data in our discussion.

The endomorphisms are associated to so-cal\-led cobordisms $(M,\partial_- M,
\partial_+ M)$. Here $M$ is a three-manifold whose boundary $\partial M$
has been decomposed in two disjoint subsets $\partial_\pm M$, each of
which can be empty. Moreover, a ribbon graph has to be chosen in $M$.
After choosing Lagrangian subspaces in $H_1(\partial_\pm M,\reals)$,
the two spaces $\partial_\pm M$ become extended surfaces. The endomorphism
associated to the cobordism is then a linear map
$$ Z(M,\partial_- M,\partial_+ M) : \quad
\calh(\partial_- M) \to \calh(\partial_+ M) \,. $$
In the application of our interest, we always choo\-se $\partial_- M$ to be 
empty. Using the fact that $\calh(\emptyset)\,{=}\,\complex\,$, we then
obtain a map 
$$ Z(M,\partial M) : \quad \complex \to \calh(\partial M) \,, $$
in other words, a line in the vector space $\calh(\partial M)$.
The image $Z(M,\partial M)1$ of the number 1 under this map then specifies
a vector in the vector space $\calh(\partial M)$ of conformal blocks.

Topological field theory thus provides a manageable way to describe explicitly
elements in the spaces of conformal blocks, a task that is very difficult
in other approaches to these spaces.

\section{2-d CFT and chiral blocks}

It is important (not only in our present context) to be aware of the fact
that the common use of the words ``\cft'' refers to two rather different 
types of physical situations. {\em Chiral conformal field theories} are 
defined on oriented manifolds $\hat X$ without boundaries. They appear, e.g., 
in the analysis of the universality class of the edge system of a 
quantum Hall sample. Indeed, the magnetic field selects a chirality and 
thereby provides an orientation of the
boundary of the sample. The main objects of chiral conformal field
theory are the spaces of conformal blocks. It is chiral CFT rather than 
full \cft\ that is the boundary theory for a topological field theory.

Full \cft\ appears in the world sheet formulation of string theory and in
the description of universality classes of critical phenomena. It can be
defined on surfaces $X$ with boundary and on unoriented surfaces. It is important
to realize that even when the surface $X$ is orientable, no orientation is 
preferred.

Chiral and full CFT are related \cite{ales,fuSc6} by a generalization of 
the mirror trick that is familiar from the treatment of boundaries in classical
electrodynamics. Given a surface $X$, one considers the double $\hat X$, a
surface that is naturally oriented. For example, the double of a disk
is the sphere, and the double of a crosscap (the real projective space
$\reals{\dl P}^2$) is a sphere as well. For general surfaces without boundary,
the double is the total space of the orientation bundle.
In all cases, there is an orientation reversing involution $\sigma{:}\;
\hat X \to\hat X$ such that $X \,{=}\, \hat X / \sigma$. 

The idea is now to construct correlators for full CFT on $X$ in terms of 
conformal blocks for the chiral CFT on its double $\hat X$ \cite{fuSc6}:
\begin{quote}
{\sl The correlators of full CFT on $X$ are specific vectors in the spaces
$\calh(\hat X)$.}
\end{quote}
The central task of constructing a full CFT from a given chiral CFT is
to specify these vectors.

These vectors must obey various consistency constraints. They encode
factorization properties as well as locality of the correlation functions 
as functions of the insertion points and of the moduli of the two-dimensional
surface.

To conclude this section, we indicate how this formulation of the problem
of constructing a full CFT is related to more conventional descriptions of
correlators. If $X$ is closed and orientable, the cover $\hat X$ consists
of two copies of $X$, but with opposite orientation. Symbolically,
$\hat X = (+X) \cup (-X)$. Correlators are blocks on $\hat X$, and hence
{\em bi\/}linear combinations of blocks on $X$.  

\section{The connecting manifold}

Combining the insights outlined in the sections 1 and 2, it is
natural to use TFT to describe the vectors in the spaces of conformal blocks
that  correspond to correlators. More precisely, for every
world sheet $X$, we construct \cite{fffs2}
a three manifold $M_X$ such that its boundary is the double,
$$ \partial M_X = \hat X \,. $$
The manifold $M_X$ will be called {\em connecting manifold\/}.
For any choice of insertion points on $X$, we will construct a Wilson
graph in $M_X$ such that 
$$ Z(M_X,\hat X): \quad \complex \to \calh(\hat X)$$
gives the correlator of the full CFT.

Let us start with two examples. When $X$ is the disk, then the double $\hat X$ 
is the sphere $S^2$. The orientation-reversing map $\sigma$ is reflection 
about the equatorial plane.
The connecting manifold $M_X$ is the full ball. Note that the intervals
perpendicular to the equatorial plane provide natural connecting lines between
the two pre-images of a bulk point, and that these connections for two different
bulk points never intersect. These connecting intervals are a general feature;
they motivate the name ``connecting manifold''. 
Our second example is the sphere, $X\,{=}\,S^2$, for which the double consists
of two disjoint copies of $S^2$. The connecting manifold is then 
the space between two concentric spheres. 

In general, $M_X$ can be
constructed as the total space of an interval bundle over the orientation
bundle, seen as a $\zet_2$-bundle. A contraction over the boundary points
ensures that $M_X$ is smooth (in this respect $M_X$ differs from
a similar manifold introduced in \cite{hora5}).

The next task is to describe the Wilson graph. In this section, we
restrict ourselves to boundary conditions that preserve all bulk symmetries.
For the topology of the disk our prescription is illustrated in the
following picture:
$$ \raisebox{-3.4em}{\scalebox{0.2}{\includegraphics{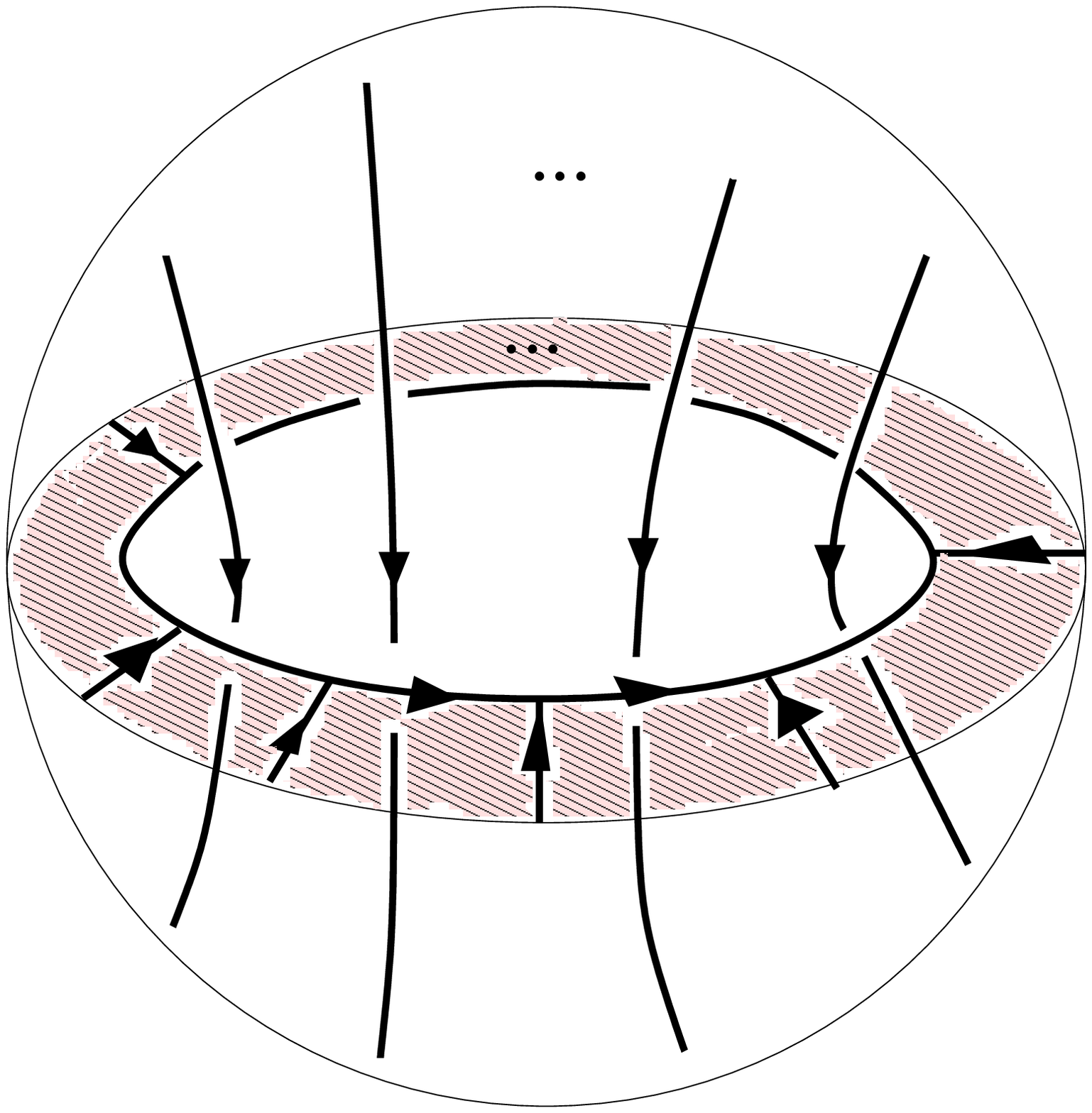}}} $$
First we join the two pre-images of a bulk point by a Wilson line along
the connecting interval. Since we work with the charge conjugation modular 
invariant, the two insertions on $\hat X$ are labelled by conjugate labels 
$\lambda,\lambda^+_{\phantom|}$, and we attach to 
the Wilson line running from the pre-image with $\lambda$ to
the preimage labelled with $\lambda^+_{\phantom|}$ the label $\lambda$. 

The components of the boundary $\partial X$ of $X$ correspond to circles
on the cover $\hat X$. (E.g.\ in the case of a disk, the single boundary
corresponds to the equator of the disk.) In the next step,
we put a circular Wilson line `close' to every such boundary circle. The 
qualification `close' needs some explanation in a topological theory: It means
here that none of the Wilson lines for bulk fields runs between the boundary
Wilson line and the boundary circle.

The boundary insertions are then joined with little Wilson lines to the
corresponding boundary Wilson line.  For each boundary insertion this 
introduces a trivalent vertex. We finally must attach labels to the boundary
part of the Wilson graph as well. For the short Wilson lines which join
the boundary insertions to the boundary Wilson line, we take the chiral 
label of the corresponding boundary field. The trivalent vertices partition
the circular Wilson line into line segments. To each such segment we
must assign a label as well. This label is interpreted as specifying a 
boundary condition. Indeed, it is known for a long time \cite{card9} that
under our hypotheses boundary conditions and primary fields are in
one-to-one correspondence.

Finally we have to deal with the trivalent vertices. One should assign
a coupling to them, i.e.\ an element in the space of three-point blocks
on the sphere. The dimension of that space is given by the fusion rules. 
Indeed, the partition function for boundary operators $\Psi^{ab}_\mu$
is nothing but the annulus amplitude, 
$$ A_{ab}(t) = \sum_\nu A_{ab}^\nu\, \chi_\nu^{}(\ii t/2) \,. $$
Under our hypotheses, the annulus multiplicities $A_{ab}^\nu$
are known \cite{card9} to be equal to the fusion rules.

We have now specified a Wilson graph in the connecting manifold $M_X$
which, according to the general rules, provides us with a specific
element in the
space of conformal blocks $\calh(\partial M) \,{=}\, \calh(\hat X)$. This
element is the correlator we are looking for:
$$ C(X) = Z(M_X,\hat X) 1 \, . $$
(In the present discussion, we have suppressed several technical details like 
the framing of the Wilson lines or the appropriate choice of La\-gran\-gi\-an
subspaces in $H_1(\hat X,\reals)$. All those details can be found in \cite{fffs3}.)

Our ansatz provides a description of correlation functions of a \cft\ in
a mathematically rigorous framework (cf.\ e.g.\ \cite{TUra}). As a consequence,
we are in a position to prove various theorems about 
correlation functions. The first statement concerns 
modular invariance. Consider the group ${\rm Aut}(\hat X,\sigma)$ of
arc-preserving homeomorphisms of $\hat X$ of degree 1 that commute
with the action of the involution $\sigma$ on $\hat X$. When $X$ is the
two-torus, this group reduces to the ordinary modular group; in the case of
surfaces with boundaries it has been called the relative modular group
\cite{bisa2}. One can show that this group acts on the spaces of conformal
blocks $\calh(\hat X)$, that the action is genuine (rather than only
projective), and that the correlators are invariant under this action.
This establishes modular invariance at all genera. 

A second collection of theorems shows that our ansatz is consistent with 
factorization both in the bulk and on the boundary. At the level of chiral
CFT, we have the following structure: Given two arcs in $\hat X$, one
can cut out little disks around these arcs and glue together the boundaries
of the two disks so as to obtain a new surface $\hat X'$ with two insertion 
points less. We label the two arcs in $\hat X$ by conjugate labels 
$\lambda,\lambda^+_{\phantom|}$ and call the corresponding labelled surface 
$\hat X_\lambda$. It follows
from the axioms of TFT that for each such gluing there is an isomorphism
at the level of spaces of conformal blocks:
$$ g_{\hat X',\hat X} :\quad \bigoplus_\lambda \calh(\hat X_\lambda) \to
\calh(\hat X') \, . $$

The correlation functions are compatible with this structure for the double. 
On a world sheet $X$ we can glue together two bulk insertions. On
the double, this amounts to a simultaneous gluing of two pairs of
insertions. For the correlation functions, one finds
$$ C(X') = \sum_\lambda S_{\lambda,\Omega}^{} \,\, g_{\hat X',\hat X}
C(X_\lambda) \,. $$
This is exactly the usual consistency constraint in full CFT, if one
takes into account the fact that in our approach
the two-point function of bulk fields is normalized to the element
$S_{\lambda,\Omega}$ of the modular matrix $S$.

Similarly, one can glue two boundary insertions. In this case, one 
deals with a single gluing on the double. One finds
$$ C(X') = \sum_\lambda (S_{\lambda,\Omega}/S_{\Omega,\Omega}) \,\,  
g_{\hat X',\hat X} C(X_\lambda) \circ \gamma \,, $$
which is again compatible with the normalization of the two-point functions
of boundary fields. (The map $\gamma$ is a natural contraction on
the space of annulus multiplicities.)

Finally one can recover the amplitudes for one bulk insertion on a disk, 
for three boundary fields on a disk, and for one bulk insertion
on the crosscap, as well as the amplitudes for annulus, Klein bottle and
M\"obius strip. Complete agreement with known results is found. 

\begin{figure}[htbp] \centering
  \scalebox{0.2}{\includegraphics{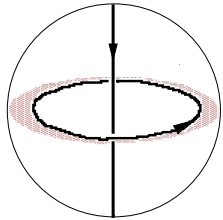}}
  \caption{$C(D_a;\lambda)$}
  \label{m4}\end{figure}

\begin{figure}[htbp] \centering
  \scalebox{0.2}{\includegraphics{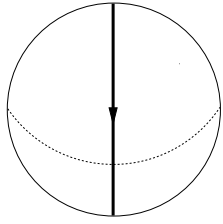}}
  \caption{$B(S^2;\lambda,\lambda^+)$}
  \label{m5}\end{figure}

As an illustrative example, we consider the case of a single bulk insertion 
$\Phi_{\lambda,\lambda^+_{\phantom I}}$ on a disk with boundary condition $a$. 
In this case the space of blocks is one-dimensional. Our task is then 
to compare the Wilson graph of figure \ref{m4} with the standard basis 
that is displayed in figure \ref{m5}. (In the present context, this 
particular conformal block $B(S^2;\lambda,\lambda^+_{\phantom I})$
is often called an `Ishibashi state'). We now obtain $S^3$
by gluing with a single three-ball. When applied to figure 
\ref{m5}, we get the unknot with label $\lambda$ in $S^3$, for which the
link invariant is $S_{0,\lambda}$. In the case of figure \ref{m4} we get a 
pair of linked Wilson lines with labels $a$ and $\lambda$ in $S^3$; the value 
of the link invariant for this graph
is $S_{a,\lambda}$. 
%

Comparison thus shows that the correlation function is
$S_{a,\lambda}/S_{0,\lambda}$ times the standard two-point block on the sphere,
$$ C(D_a;\lambda) = (S_{a,\lambda}/S_{0,\lambda}) \cdot 
B(S^2;\lambda,\lambda^+_{\phantom|}) \,.  $$ 
Taking into account the normalization of bulk fields, we recover the known 
result that the correlator for a canonically normalized bulk field $\lambda$
on a disk with boundary condition $a$
is $S_{a,\lambda}/\!\sqrt{S_{0,\lambda}}$ times the standard two-point block
on the sphere. (This relation forms the basis of the so-called boundary
state formalism \cite{card9}.)

\section{Symmetry breaking boundary conditions}

We now study the more general case of boundary conditions that break part of the
bulk symmetries: We only require that some subalgebra $\calap$ of the algebra 
$\cala$ of chiral symmetries is preserved by the boundary conditions.

A particularly simple realization of this situation arises when $\calap$
is an {\em orbifold subalgebra} of $\cala$: Fix a group $G$ of automorphisms
of $\cala$ and define $\calap$ to be the subalgebra of elements of $\cala$ 
that are left pointwise fixed by all automorphisms in $G$. In case $G$
is a finite abelian group, many aspects of boundary conditions that preserve
only $\calap$ are known \cite{fuSc1112}. The two most important insights in 
our present context are the following:
\nxt Bulk and boundary fields carry labels from two distinct sets.
\nxt Boundary conditions that break bulk symmetries correspond
to {\em twisted representations\/} of the chiral algebra $\cala$. \\
There is a notion of fusion for such representations \cite{gabe7}, 
and the annulus coefficients can be expected \cite{fuSc10,bppz2}
to coincide with the fusion rules of twisted representations.

Together with the observation that in our Wilson graphs bulk and boundary 
fields always lie in different connected components of a graph, 
the first point suggests the description in terms of a
new fusion ring for the boundary fields.
We stress that this fusion ring cannot be expected to be
modular. Indeed, for twisted representations the conformal weight does 
not have the same fractional part for all states in the module, so there
is no `twist'. Since every braided tensor $*$-category with conjugates 
automatically has a twist, one cannot expect a braiding either.

We now concentrate on the boundary fusion ring. Its structure 
constants are the annulus multiplicities. It can be obtained from
the fusion ring of the $\calap$-theory by the following general recipe
\cite{fuSc14}:\,%
 \footnote{~This recipe can be put on a firmer mathematical basis in thoses
 cases where a description of the CFT in terms of nets of factors is known.
 In this case, it amounts to $\alpha$-induction; for a review and
 references see \cite{boev5}. For our present purposes, it is sufficient
 to consider the structure at the level of TFT only.}
The vacuum module $\calh_\Omega$ of the $\cala$-theory decomposes
into $\calap$-modules with certain multiplicities:
$$ \calh_\Omega  = \bigoplus_{\bar\mu} n_{\bar\mu} \bar\calh_{\bar\mu} \,. $$
The non-negative integers $n_{\bar\mu}$ define an element $\bar\theta$ of the 
fusion ring of $\calap$:
$$ \bar\theta = \sum_{\bar\mu} n_{\bar\mu} \Phi_{\bar\mu} \,. $$

To every element $\Phi_{\bar\lambda}$ of the fusion ring for $\calap$ one now
associates an element $\alpha_{\bar\lambda}$ in a new fusion ring $\cala'$.
This operation preserves multiplication, addition and conjugates:
$$ \alpha_{\bar\lambda}^{} {\star}\,\alpha_{\bar\mu}^{} \,{=}\,
\alpha_{\bar\lambda\star\bar\mu}^{} \,, \ \ \alpha_{\bar\lambda}^{} 
{+}\,\alpha_{\bar\mu}^{} \,{=}\, \alpha_{\bar\lambda+\bar\mu}^{}
\,, \ \ {(\alpha_{\bar\lambda})}^+_{} \,{=}\, \alpha_{\bar\lambda}^+ $$
This would not lead to anything new, were it not for another
definition, namely of the spaces of homomorphisms: We set
$$ \Hom_{\cala'}^{}(\alpha_{\bar\lambda}^{},\alpha_{\bar\mu}^{}) :=
\Hom_{\calap}^{}(\Phi_{\bar\lambda},\bar\theta\star\Phi_{\bar\mu}) \, . $$

This relation implies that even when $\bar\lambda$ is simple, i.e.\ corresponds 
to an irreducible representation of the chiral algebra $\calap$, the induced
object $\alpha_{\bar\lambda}$ is not necessarily simple. It may even happen
that the category $\cala'$ does not contain enough simple objects to
decompose every object into a direct sum of simple objects. This is
a generalization of the problem of fixed point resolution in 
simple current extensions. A general construction in tensor $*$-categories
guarantees the existence of a bigger category $\cala_{\bound}$ in which
the fixed points are resolved. Unfortunately, this prescription cannot
(yet) be made as explicit as \cite{fusS6} in the simple current case.

We take this resolved tensor category $\cala_{\bound}$ as the boundary category.
Its simple objects correspond to (elementary) boundary conditions.
The structure constants in the tensor product correspond to the annulus
multiplicities. The sectors actually come in two classes, solitonic or
local. Local sectors correspond to symmetry preserving boundary conditions.
The solitonic sectors correspond, in the case of simple current extensions, 
to boundary conditions with non-va\-ni\-shing monodromy charge, which implies
a non-trivial automorphism type (or, equivalently, a non-trivial gluing 
automorphism on the boundary).  More generally, symmetry breaking boundary
conditions are in one-to-one correspondence to solitonic representations of 
the chiral symmetry $\cala$.
The resolved tensor category $\cala_{\bound}$ is associative and
closed under subobjects. Thus one can define $6j$-symbols; the same arguments
as in the Cardy case \cite{bppz2,fffs} then show that the OPE of two boundary
fields coincides with these $6j$-symbols.\,%
 \footnote{~Note that in our approach all 6 indices of the $6j$-symbols
 take their values in the same label set, namely the one for the
 simple objects of the boundary category $\cala_{\rm bound}$,
 just like all 3 indices of the annulus coefficients do.
 For a different approach to symmetry breaking boundary conditions
 we refer to \cite{bppz2} and Zuber's contribution to these proceedings.
 In that setting, the boundary OPE is described by
 generalized fusing matrices with two types of indices, and likewise the
 annulus coefficients carry two types of indices, one of them being a
 label for simple objects of the $\bar{\cala}$-theory.}

To conclude this contribution, let us emphasize that the space of boundary
conditions carries a surprising amount of beautiful structure. The 
presence of this structure makes us confident that many more problems can be 
tackled than one could have hoped for some time ago.  

\vskip2em
\noindent
{\small
{\sc Acknowledgement}:\\
The results presented in Section 3 have been obtained in 
collaboration with Giovanni Felder and J\"urg Fr\"ohlich. We would like to
thank them for a very pleasant collaboration as well as for discussions
on the contents of Section 4.}

  \newcommand\J[5]   {{#1} {\bf #2}, {#4} ({#3})}
  \newcommand\Prep[2]{preprint {#1}}
  \newcommand\inBO[7]{in:\ {\em #1} ({#3}, {#4} {#5}), p.\ {#6}}
 \newcommand\wb{\,\linebreak[0]} \def\wB {$\,$\wb}
 \newcommand\Bi[1]    {\bibitem{#1}}
 \newcommand\BOOK[4]  {{\em #1\/} ({#2}, {#3} {#4})}
 \def\jf    {J.\ Fuchs}
 \def\bams  {Bull.\wb Amer.\wb Math.\wb Soc.}
 \def\comp  {Com\-mun.\wb Math.\wb Phys.}
 \def\ijmp  {Int.\wb J.\wb Mod.\wb Phys.\ A}
 \def\jgap  {J.\wb Geom.\wB and\wB Phys.}
 \def\jopa  {J.\wb Phys.\ A}
 \def\nuci  {Nuovo\wB Cim.}
 \def\nupb  {Nucl.\wb Phys.\ B}
 \def\phlb  {Phys.\wb Lett.\ B}
 \def\phrl  {Phys.\wb Rev.\wb Lett.}
 \newcommand\fscp[2] {\inBO{Fields, Strings, and Critical Phenomena} {E.\
     Br\'ezin and J.\ Zinn-Justin, eds.} \NH{Amsterdam}{1989} {{#1}}{{#2}}}
 \def\Ca     {{Cambridge}}
 \def\CUP    {{Cambridge University Press}}
 \def\NH     {{North Holland}}
 \def\NY     {{New York}}
 \def\PL     {{Plenum Press}}

\end{document}